\documentclass[12pt,preprint]{aastex}

\def\Gb{\Gamma}

\def\pf{p_f}
\def\pr{p_r}

\def\g43{\gamma_{43}}
\def\b43{\beta_{43}}
\def\tobs{t_{\rm obs}}

\def\Gej{\Gamma_{\rm ej}}
\def\Ee{E_{\rm ej}}
\def\Ehead{E_{\rm head}}
\def\Etail{E_{\rm tail}}

\def\rhoej{\rho_{\rm ej}}
\def\nejRS{n_{\rm ej}{\rm (RS)}}
\def\GejRS{\Gamma_{\rm ej}{\rm (RS)}}

\def\tobs{t_{\rm obs}}

\def\Gend{\Gamma_{\rm end}}
\def\thjet{\theta_{\rm jet}}
\def\M0{M_0}

\def\be{\begin{equation}}
\def\ee{\end{equation}}
\def\beq{\begin{eqnarray}}
\def\eeq{\end{eqnarray}}

\newbox\grsign \setbox\grsign=\hbox{$>$} \newdimen\grdimen \grdimen=\ht\grsign
\newbox\simlessbox \newbox\simgreatbox \newbox\simpropbox
\setbox\simgreatbox=\hbox{\raise.5ex\hbox{$>$}\llap
     {\lower.5ex\hbox{$\sim$}}}\ht1=\grdimen\dp1=0pt
\setbox\simlessbox=\hbox{\raise.5ex\hbox{$<$}\llap
     {\lower.5ex\hbox{$\sim$}}}\ht2=\grdimen\dp2=0pt
\setbox\simpropbox=\hbox{\raise.5ex\hbox{$\propto$}\llap
     {\lower.5ex\hbox{$\sim$}}}\ht2=\grdimen\dp2=0pt

\def\simlt{\mathrel{\copy\simlessbox}}

\begin{document}

\title{On the Mechanism of Gamma-Ray Burst Afterglows}

\author{Z. Lucas Uhm, Andrei M. Beloborodov\altaffilmark{1}}
\affil{Physics Department and Columbia Astrophysics Laboratory, \\
       Columbia University, 538 West 120th Street, New York, NY 10027.}

\altaffiltext{1}{Also at Astro-Space Center, Lebedev Physical Institute,
Profsojuznaja 84/32, Moscow 117810, Russia}

\begin{abstract}
The standard model of afterglow production by the forward shock wave
is not supported by recent observations.
We propose a model in which the forward shock 
is invisible and afterglow is emitted by 
a long-lived reverse shock in the burst ejecta.
It explains observed optical and X-ray light curves, including
the plateau at $10^3-10^4$~s with a peculiar chromatic break, and the second 
break that was previously associated with a beaming angle of the explosion.
The plateau forms following a temporary drop of the reverse-shock pressure much below 
the forward-shock pressure.
 A simplest formalism that can describe 
  such blast waves is the ``mechanical'' model 
  (Beloborodov \& Uhm 2006); we use it in our calculations.
\end{abstract}



\section{INTRODUCTION}

Since its discovery, gamma-ray burst (GRB) afterglow has been 
attributed to the forward shock wave (M\'esz\'aros \& Rees 1997). 
This hypothesis was consistent with first follow-up observations at 
$\tobs\sim 0.1-10$~days after prompt GRBs. However, recent observations of 
early afterglows ($\tobs< 0.1$~day) 
are difficult to reconcile with the model (e.g. Zhang 2007).
One outstanding problem is the low-level plateau in the X-ray light curve
(Nousek et al. 2006); this and other 
problems are summarized in \S~3, where we discuss their possible resolution.

Besides the forward shock (FS), a reverse shock (RS) is expected 
in the burst ejecta. A short-lived RS was proposed to emit a brief optical 
flash (M\'esz\'aros \& Rees 1999; Sari \& Piran 1999), and later absence 
of such flashes was interpreted as a lack of RS emission (Roming et al. 2006).
A long-lived RS in stratified ejecta with a decreasing Lorentz factor $\Gej$ 
was proposed as a mechanism of gradual energy injection into the FS 
(Rees \& M\'esz\'aros 1998) and the RS emission
was invoked to explain radio data (Panaitescu \& Kumar 2004).

The interaction of power-law ejecta with a power-law medium obeys
scaling laws that are derived analytically (Rees \& M\'esz\'aros 1998). 
However, the general problem of explosions driven by ejecta with arbitrary 
stratification of $\Gej$ remained unsolved. We have developed a simple 
``mechanical'' formalism for such explosions (Beloborodov \& Uhm 2006; 
hereafter BU06), which allows us to explore a new class of dynamical models, 
with rapid and strong evolution of the RS.
We calculate the emissions from FS and RS 
and search for a scenario that would reproduce the observed 
X-ray and optical light curves. 
We find that observations can be explained if it is only the RS that contributes
to the observed afterglow and the FS is invisible because of
its extremely low radiative efficiency.


\section{AFTERGLOW MODEL}

The central explosion ejects a cold (adiabatically cooled) relativistic flow.
It may be viewed as a sequence of shells of energy $\delta\Ee$ 
that coast with Lorentz factors $\Gej$. Each shell can be prescribed an 
ejection time $\tau$ at a small radius $r$, and functions $d\Ee/d\tau$ 
and $\Gej(\tau)$ completely describe the ejected flow. 
The ejecta density is derived from continuity equation,
\be
  \rhoej(\tau,r)=\frac{d\Ee/d\tau}{4\pi r^2\Gej^2 
                  c^3\left(1-r\Gej^\prime/c\Gej^3\right)},  
\label{rhoej}
\ee  
where $\Gej^\prime\equiv d\Gej/d\tau \leq 0$.
The ejecta push the blast --- a thin shell of compressed gas between 
FS and RS --- as described by equations~(11)-(16) in BU06. 
We solve these equations numerically, tracking self-consistently the 
location of RS in the ejecta. The blast-wave model is
determined by three functions: $\rho_1(r)$ (external density),
$\Gej(\tau)$, and $\Ee(\tau)$. 
When the condition $|d\Gej/d\tau| \gg c\Gej^3/r$ is satisfied,
equation~(\ref{rhoej}) simplifies,
\be
\label{eq:app}
  \rhoej=\frac{\Gej}{4\pi r^3c^2}\,\left|\frac{d\Gej}{d\Ee}\right|^{-1},  
   \qquad
  \frac{r}{c\Gej^2}\gg\tau\,\left|\frac{d\log\Gej}{d\log\tau}\right|^{-1}.
\ee
Then the ejection time $\tau$ drops out, and
only one function $\Gej(\Ee)$ describes the ejecta.

Once the dynamical equations are solved, we calculate the
synchrotron emission from the shocked medium (FS emission) and shocked 
ejecta (RS emission). This calculation uses the standard model with 
three parameters (e.g. Piran 2004): fraction $\epsilon_e$ of the shock 
energy that goes to electron acceleration, slope $p$ of the 
electron spectrum, and magnetic 
parameter $\epsilon_B=(B^2/8\pi U)<1$, where $U$ is
the energy density of the shocked gas.
The blast is treated as one body in the sense that it moves as one body, 
with a common $\Gb$. However, we take into account that it is made 
of different hot shells that pile up from the FS and RS. 
The synchrotron emissivity of each shell is tracked as the blast expands.
The sum of observed emissions from all shells 
is found taking into account the velocity and curvature of the shells.

The main points of this work are demonstrated by the numerical model
presented in Figures~1-2. It assumes a uniform external medium 
with $n_1(r)=\rho_1/m_p=1$~cm$^{-3}$ and a total isotropic energy of 
the explosion 
$E_b=10^{54}$~ergs. The stratification function $\Gej(\Ee)$ 
(Fig.~1) assumes a ``head'' of the ejecta with $\Gej=300$, which carries 
an energy $\sim E_b/3$. The remaining 2/3 of energy is carried by 
a slower ``tail.''
The RS crosses the head at early times $\tobs\simlt 10^2$~s and proceeds
to the tail. Equation~(2) becomes valid in the tail.

\begin{figure}
\begin{center}
\plotone{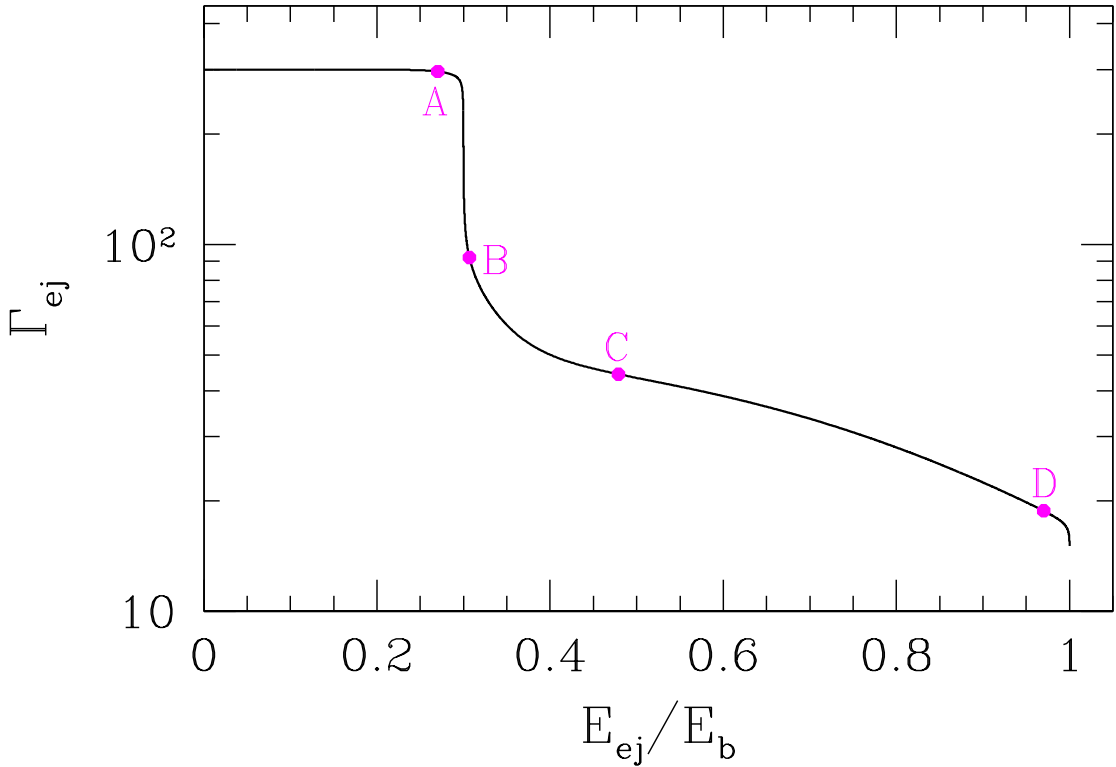}
\caption{\small
Stratification function $\Gej(\Ee)$ of the numerical model.
Here $\Ee(\tau)$ is kinetic energy of the ejected flow. It serves 
as a Lagrangian coordinate: $\Ee=0$ corresponds to the first ejected 
shell and $\Ee=E_b$ corresponds to the last shell; $E_b=10^{54}$~ergs 
is the total energy of explosion.
The reverse shock (RS) starts at $\Ee=0$ and moves toward $\Ee=E_b$, 
passing through points A, B, C, and D.
The transition from ``head'' to ``tail'' of the ejecta occurs
between points A and B. Point C is close to the inflection point of
function $\Gej(\Ee)$ and corresponds to the maximum of $\rhoej$ (Fig.~2).
Point D is near the end of the ejecta.
In the exact numerical model, the ejected flow is described by two functions, 
$\Gej(\tau)$ and $\Ee(\tau)$ with $0<\tau<\tau_b\approx 100$~s. However,
for the afterglow emitted after the RS passes 
point A, only one function is important --- $\Gej(\Ee)$ 
--- and $\tau\approx 0$ can be assumed for all shells after point A
(see the text and eq.~\ref{eq:app}). Therefore, we show only $\Gej(\Ee)$
here.
}
\end{center}
\end{figure}

The emission produced by the FS with usual 
parameters $\epsilon_e=0.1$, $\epsilon_B=0.01$, and $p=2.3$ is
shown by the thin curves in Figure~2e. Both X-ray and optical light curves
are inconsistent with observations and illustrate the problems
of the FS model. The theoretical 
optical light curve peaks at $10^3-10^4$~s, when the peak frequency 
of synchrotron emission passes through 
the optical band (e.g. Sari et al. 1998). 
This peak is not observed. The theoretical 
X-ray light curve has a long monotonic 
decay with slope $\alpha\sim 1$.\footnote{It deviates from the power-law 
because of energy injection into the FS from the tail. The growth of 
blast-wave energy (by a factor of 3) implies a deviation from the standard 
deceleration law $\Gb^2\rho_1r^3=const$.} 
By contrast, observed X-ray afterglows are much weaker at $10^2-10^4$~s; 
they show an initial steep decay to a low emission level and then a plateau. 

\begin{figure}
\begin{center}
\plotone{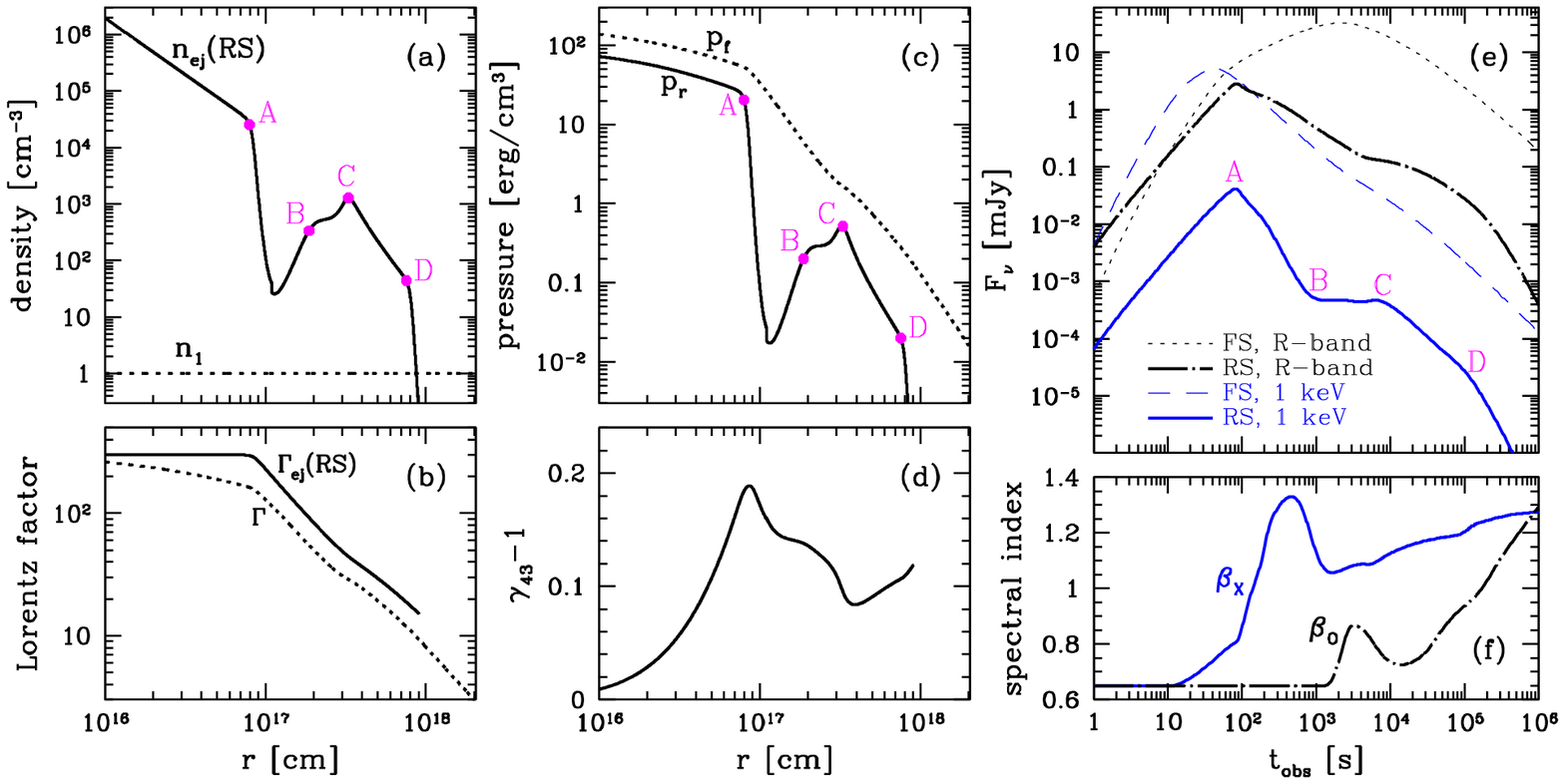}
\caption{\small
Panels (a)-(d) show the blast-wave evolution in the mechanical model.
(a) Density of the preshock ejecta $\nejRS=\rhoej/m_p$ as a function
of radius $r$ of the expanding blast wave; the assumed external density 
$n_1=\rho_1/m_p=1$~cm$^{-3}$ is shown for comparison.
(b) Lorentz factors of the preshock ejecta, $\GejRS$, and the blast, $\Gamma$.
(c) Pressures at the FS and RS, $\pf$ and $\pr$.
(d) Relative Lorentz factor of the RS, $\gamma_{43}$ (see BU06).
Panels (e) and (f) show observed emission from the blast wave as
a function of observer time. The emission parameters are $\epsilon_B=0.01$,
$\epsilon_e=0.1$, and $p=2.3$. (e) Observed spectral flux, 
assuming the burst is at a cosmological redshift $z=1$. 
Thin curves show the FS emission
--- X-ray (dashed line) and optical (dotted line). Thick curves show the RS emission 
--- X-ray (solid line) and optical (dash-dotted line).
(f) Evolution of spectral indices $\beta_X$ (X-ray) and $\beta_O$ (optical) 
for the RS emission. The transition from the slow cooling 
spectrum $\beta=(p-1)/2=0.65$ ($\nu_c>\nu$) to the fast cooling 
spectrum $\beta=p/2=1.15$ ($\nu_c<\nu$) occurs early in the X-ray band 
and later in the optical band. The peaks in $\beta_X$ and $\beta_O$ will be 
discussed elsewhere (Z.~L.~Uhm \& A.~M.~Beloborodov 2007, in preparation).
The correspondence between the marked points in dynamical panels (a)-(d) 
and emission panels (e)-(f) is approximate:
 radiation received at a given $\tobs$ is emitted by the blast 
 as it propagates a range of radii $r\sim 2\Gb^2\tobs c$.
}
\end{center}
\end{figure}

The RS emission for the same explosion is shown by thick curves
in Figure~2e. The same parameters $\epsilon_e=0.1$, $\epsilon_B=0.01$,
and $p=2.3$ are assumed for the RS. Then the cooling frequency $\nu_c$ 
is between optical and X-ray bands throughout most of the afterglow.

The X-ray emission is produced by fast-cooling electrons near the RS.
Its initial peak (marked A in Fig.~2e) is followed by 
the steep decay AB and the plateau BC. This behavior can be understood 
from Figures 2a-2d, which show the explosion dynamics. 
The ejecta density drops dramatically behind the head of the ejecta
because of the large gradient of $\Gej$ (see Fig.~1 and eq.~\ref{eq:app}).
Therefore, X-ray emission is strongly reduced when the RS enters the tail.
Its decay at $\tobs=10^2-10^3$~s has a temporal index $\alpha\sim 3$ and 
is limited only by the spherical curvature of the RS.
The X-ray emission does not recover until the RS propagates to the 
region of flatter stratification function (smaller $|d\Gej/d\Ee|$ and
higher $\rhoej$).
This recovery begins at $\tobs\sim 10^3$~s and corresponds to the beginning 
of the X-ray plateau (point B). During the plateau stage, $\rhoej$ at the 
RS grows and reaches a maximum at point C where $|d\Gej/d\Ee|$ is near its 
minimum. This point is the end of the X-ray plateau.

Following point C, the X-ray light curve has a slope $\alpha\sim 1$ 
until the last
break at point D. This break corresponds to the RS reaching 
the end of the ejecta, and its observed time is $\tobs\sim r/2\Gend^2c$.
In our example model, $\Gend\approx 20$ (Fig.~1), which gives the break
D at $\tobs\sim 10^5$~s. 
(We also ran models with smaller $\Gend$; then the power-law with 
$\alpha\sim 1$ extends to longer times $\sim$ weeks.) The 
steep slope of the light curve after break D
is limited by the spherical curvature of the RS. 

The RS optical light curve also differs significantly from the prediction 
of the FS model. It has an initial power-law decay at $10^2-3\times10^3$~s 
followed by the hump or ``shoulder.'' The optical light curve can be 
understood by noting that it is produced by slow-cooling electrons. 
When crossing the head of the ejecta, the RS creates a population of 
nonthermal electrons that continue to emit optical radiation after point A, 
when the RS weakens dramatically.
This residual emission has a power-law light curve with a slope 
$\alpha\sim 1$. It is approximately described by the formula 
(see Beloborodov 2005b, \S~7),
\be
\label{eq:al}
   \alpha=\frac{p+1}{16}+\frac{5p+1}{8\hat{\gamma}},
\ee
where $\hat{\gamma}$ is the adiabatic index of the shocked ejecta.
The RS is only mildly relativistic, and adiabatic cooling
quickly reduces the sound speed to a non-relativistic value, 
so $\hat{\gamma}\approx 5/3$. 
Equation~(\ref{eq:al}) then gives $\alpha\approx 1.1$ for $p=2.3$.

The RS recovery near point C makes an additional contribution to optical 
emission and produces the shoulder in the light curve. In contrast to the 
X-ray afterglow, it does not break at point C.
We ran a number of explosion models with different $\Gej(\Ee)$ and 
found that the optical shoulder usually accompanies the X-ray plateau. 
This feature is less prominent if $\Etail<\Ehead$ (then it is buried 
by the residual optical emission from the head).
For high $\epsilon_B\sim 0.1$, 
  the optical-emitting electrons enter the fast-cooling regime
before point C and the optical light curve has a plateau with a break
at C, similar to that in the X-ray band.

Figure~2f shows spectral indices $\beta_O$ and $\beta_X$ in the optical 
and X-ray (1~keV) bands for the RS emission.
The X-ray break at the end of the plateau is not accompanied
by any significant change of $\beta_X$.
This is consistent with observations (O'Brien et al. 2006). 

We ran the same model but with three bumps in external density $n_1(r)$ 
(Fig.~3). Figure~3d compares the afterglow with and without the bumps.
FS emission is barely changed (this conclusion is similar to that of 
Nakar \& Granot 2006), and RS emission is changed significantly. 
The bump decelerates the blast, which weakens the 
FS and makes the RS stronger ($\gamma_{43}$ temporarily increases;
see Fig.~3b). 
The electron spectrum injected at the RS is then shifted to higher 
energies $\propto(\gamma_{43}-1)$ and its synchrotron emission increases.
The RS re-brightening is especially large if the bump occurs
when the blast is near point C.
The mechanical model does not take into account sound waves or shock waves 
that may be generated in the blast when it hits a bump, so the 
dynamics shown in Figure~3 is approximate. Nevertheless, this calculation 
well illustrates the difference between FS and RS. 

\begin{figure}
\begin{center}
\plotone{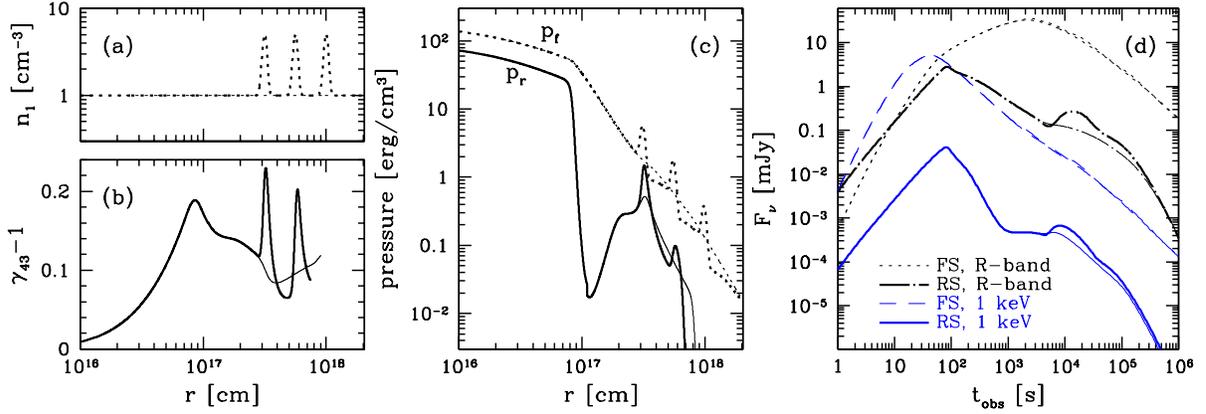}
\caption{\small Same explosion model as in Fig.~2, but with three density
bumps in the external density $n_1(r)$. Panel (a) shows the positions 
and amplitudes of the bumps. (b) Evolution of the RS relative Lorentz factor, 
$\gamma_{43}$. (c) Evolution of pressures in the FS and RS, $\pf$ and
$\pr$. (d) Same as Fig.~2e, but for the model with bumps. 
For comparison, the model without bumps (from Fig.~2) is shown 
by thin curves in all panels.
}
\end{center}
\end{figure}


\section{DISCUSSION}

If the FS dominates afterglow production, several puzzles arise. 
We list these problems below and discuss how they are avoided/solved 
in the proposed RS model.

1. The FS model predicts much stronger emission at $10^2-10^3$~s 
than observed. The FS does not stop producing the early afterglow --- 
it cannot weaken until $\Gb$ is reduced and $\tobs\sim r/\Gb^2c$ becomes
large --- while data suggest that the X-ray afterglow is temporarily 
suppressed at early times $\tobs\sim 10^2-10^3$~s. 

The RS weakens abruptly at $\tobs\sim 10^2$~s,
when it crosses the head of the ejecta.
This weakening explains the observed ``lack'' of early X-ray afterglow.

2. The FS model does not easily explain the shallow decay (plateau) in 
the X-ray light curve at $\tobs\sim 10^3-10^4$~s (Nousek et al. 2006; 
Zhang 2007; Granot 2006). Its predicted luminosity in the fast-cooling 
regime is $L\sim\epsilon_e E/\tobs$, where $E$ is the blast-wave energy. 
It can give a flat light curve if $E(\tobs)$ grows, i.e. if energy is 
gradually injected into the blast wave from the ejecta tail. 
However, this explanation invokes a huge energy injection (e.g.
an increase of $E$ by a factor of 3 does not help, cf. Fig.~2). 
A plateau of 1-2 orders in $\tobs$ would require a tail $\sim 100$ times 
more energetic than the head of ejecta. 
On the other hand, the head should emit the prompt $\gamma$-ray burst
that strongly dominates the observed energy output. This would require 
an extremely high radiative efficiency for the prompt emission. 

The RS produces the plateau BC as it recovers after the steep transition 
to the tail of the ejecta and enters the shallow part of the stratification 
function $\Gej(\Ee)$ (Figs.~1 and 2a). 
In this scenario, $E_{\rm tail}\sim E_{\rm head}$ and $1-10$\% efficiency
of the prompt emission is sufficient. 

3. The observed peculiar break at the end of the plateau is difficult to
reconcile with the FS model. The end of energy injection in the model
causes a steepening of $\Gb(r)$, and the FS emission must respond
to this steepening in all bands, both fast-cooling and slow-cooling.
By contrast, the observed break often shows up only in X-rays
(Panaitescu et al. 2006). Such chromatic breaks could appear when  
the X-ray spectrum steepens ($\nu_c$ passes through the XRT band), 
but this is excluded: the X-ray spectrum does not change across the break.

This break is reproduced by the RS model (point C in Figs.~1-2),
with no abrupt changes in $\beta_X$. 
Emission in slow-cooling bands does not show a break;
instead, a shoulder in the light curve is predicted (\S~2).

4. To explain the second break that was observed in many afterglows at 
$\tobs\sim 10^5$~s, the FS model assumes a small beaming angle of the 
ejecta, $\thjet$. When $\Gb$ decreases below $\thjet^{-1}$ the 
``jet break'' must inevitably appear in all bands (Rhoads 1999).
By contrast, the observed breaks at $\sim 10^5$~s are often 
chromatic or, in some bursts, not seen at all (e.g. Zhang 2007).

The RS model predicts the second break when the shock reaches the end of 
ejecta or faces a steep decline in $\rhoej$ (point D in Figs.~1-2). 
Its observed time is $\sim 1(\Gend/20)^{-2}$~days.
Afterglow steepens immediately only in fast-cooling bands.\footnote{
The light-curve slope becomes $\alpha=2+\beta$ if the RS weakens sharply 
at D --- the slope is then controlled by photons emitted at moment D 
and arriving with a delay because of the spherical curvature of the RS. 
By contrast, the usual jet model predicts $\alpha=p$ after the break.}
No dramatic change occurs in the blast deceleration at this point, 
and the slowly-cooling emission steepens with a delay.

5. The FS model does not easily explain re-brightenings observed in 
some optical afterglows. The response of FS emission to bumps 
in external density is weak (see Fig.~3 and Nakar \& Granot 2006), 
while observed re-brightenings are strong, up to one order in flux. 

The RS emission is sensitive to external bumps (Fig.~3).
Inhomogeneities in the ejecta may also cause bumps in the light curve.

All five problems of the FS model ultimately have one common
reason: changes in emission must invoke significant changes in $\Gb$. 
By contrast, the RS model does not require strong deviations of $\Gb(r)$ 
from a power-law,
and the afterglow light curves are shaped by the ejecta stratification.
As a result, the temporal properties of the RS emission are 
consistent with observations, while those of the FS are not.

As a solution, we propose that the entire afterglow is produced by the RS.
The FS may be invisible for two reasons:
(1) Magnetic fields are too weak in the external medium and not 
sufficiently amplified by the FS. If $\epsilon_B\sim 10^{-7}$,
the FS emission is negligible.
(2) Electrons are not efficiently accelerated in the FS
(e.g. diffusive acceleration is inefficient
in ultra-relativistic shocks; see Beloborodov 2005b).
By contrast, the ejecta may carry strong magnetic fields.
Besides, the RS is mildly relativistic,
which makes electron acceleration more plausible. 
This leaves the RS as the main producer of afterglow,
and consistency with data is achieved.

This model has interesting implications. 
The mean energy per electron in the RS is much lower than it would be 
in the FS model (while the number of electrons is much larger, and
the dissipated energy in the RS is not much below that in the FS). 
Emission from most electrons is self-absorbed, and only a tail of 
their spectrum contributes to the X-ray and optical afterglow.
The net radiative output of the blast wave peaks at the cooling frequency 
$\nu_c$; however, an energetic fraction $\sim (\nu/\nu_c)^{(3-p)/2}$ is 
emitted at low $\nu\ll\nu_c$. The RS afterglow should
be very bright in the infrared bands, and 
this prediction may be tested by observations.

If afterglow is indeed produced by the shocked ejecta rather than the 
shocked external medium, it carries important information about the 
explosion. We find that GRB ejecta are made of two blocks --- head 
and tail --- with different $\Gej$ and comparable kinetic energy.
Its magnetic field is well below equipartition, $\epsilon_B\ll 1$.
The RS emission is sensitive to the stratification function 
$\Gej(\Ee)$, which explains the diversity of afterglows.
We have shown here one example with a flat plateau, and other examples
will be  presented in an accompanying paper.
The plateau has a larger slope $\alpha$ (and a longer duration) if
$\Gej(\Ee)$ is less concave at the beginning of the tail.
The ``steep decay + plateau'' shape disappears if $\Gej$ decreases
gradually after point A instead of jumping from $\sim 300$ to $<100$.
In our models, the plateau often ends with a small flare,  
which is also observed (O'Brien et al. 2006).

The study of radio emission expected in the RS model will be published 
elsewhere. Self-absorption limits the peak of radio afterglow, and 
observed light curves provide additional constraints on the model. 
The inverse Compton emission of high-energy $\gamma$-rays will also be 
investigated elsewhere. At least two high-energy components are 
expected: a brief flash that is produced through upscattering of prompt 
$\gamma$-rays by the electrons accelerated at the RS (Beloborodov 2005a) 
and the usual synchrotron self-Compton emission that can extend to 
longer times.  

\acknowledgements
This work was supported by NASA 
Swift grant, cycles 2 and 3.



\end{document}